\documentclass[aps,prl,preprint,groupedaddress]{revtex4}

\begin{document}


\title{Some exact stationary state solutions of a nonlinear Dirac equation in 2+1 dimensions}


\author{Patrick Das Gupta}
\email[]{patrick@srb.org.in}
\author {Samiran Raj and Debapriya Chaudhuri}
\affiliation{Department of Physics and Astrophysics, University of Delhi, Delhi - 110 007 (India)}


\date{\today}

\begin{abstract}

Graphene's honeycomb lattice structure is quite remarkable in the sense that it leads,  in
the long wavelength limit, to a massless Dirac equation description of nonrelativistic
quasiparticles associated with  electrons and holes present in the two dimensional crystallite. In the case of cold bosonic atoms
 trapped in a honeycomb optical lattice, Haddad and Carr (2009) have recently shown, by taking into account binary contact
 interactions, that the dynamics of these Bose-Einstein condensates is governed by a nonlinear Dirac equation (NLDE). In this paper, we study 
 exact stationary solutions of 
such a NLDE. After proving that the energy eigenvalues are real, we show that the sum of  orbital angular momentum and 
 pseudospin angular momentum normal to the crystal commutes with the nonlinear Hamiltonian whenever magnitudes of the pseudospin
 components 
 do not depend on the polar angle $\phi $. We obtain some exact stationary and localized solutions of the NLDE.
\end{abstract}

\pacs{}

\maketitle

\section{I. INTRODUCTION}
Graphene, peeled off from bulk graphite by means of micromechanical cleavage technique, is a zero-gap semiconductor [1,2]. 
Being a monolayer of carbon atoms arranged in a planar hexagonal lattice structure in which charge carriers travel
distances that are thousand times the lattice spacing without being subjected to any scattering, graphene helped in
establishing that stable two dimensional crystals do exist, contrary to what one expects from thermodynamic instability arguments
put forward by Landau,
 Peierls and Mermin in the past [3,4].
 
For wavelengths  much\
larger than the lattice size,  charged quasiparticles  in the honeycomb lattice  satisfy a massless Dirac equation [5-7], 
$$ i \hbar {{\partial \psi}\over{\partial t}}  = - i \hbar v_F (\vec \sigma . \nabla ) \psi  \eqno(1)$$
where,
\[
\psi =
\left( {\begin{array}{cc}
 \psi _A  \\
 \psi _B  \\
 \end{array} } \right)
\]
$$\nabla = \hat i \frac {\partial} {\partial x} + \hat j \frac {\partial} {\partial y}$$
$$\vec \sigma=\hat i \sigma_x + \hat j \sigma_y$$
and $v_F\approx 0.01 \ c$ represents the Fermi speed.
The components $\psi_A$ and $\psi_B$ are  bi-component spinors describing spin-half charge carriers, while subscripts A and B  refer to the two
 inequivalent sites A and B of the sublattice forming the hexagonal structure, representing
 pseudospin degrees of freedom. The associated Clifford algebra,
$$\gamma^\mu \gamma^\nu +   \gamma^\nu \gamma^\mu = 2\ \eta^{\mu \nu}, \ \mu , \nu = 0,1,2 $$
is satisfied  by,
$$\gamma^0 = \sigma_z  , \gamma^1 = \sigma_z \sigma_x, \gamma^2 = \sigma_z \sigma_y  \ .$$
 Relativistic quantum features  like {\it zitterbewegung} and Klein paradox, which so far have been confined to the domain of quantum electrodynamics,
 make their appearance in the context of subrelativistic charge carriers in graphene because of eq.(1) [8].

In an interesting recent paper, Haddard and Carr have studied weakly interacting bosonic atoms anchored to a planar honeycomb optical lattice [9].
The quantum many body theory of this system is described by a Hamiltonian,
$$\hat H = \int { }\int {\Phi ^\dagger \bigg (\frac {-\hbar^2} {2m} \nabla^2 + V(\vec r) \bigg )\Phi \ dx \ dy} + \lambda \int { }\int{\Phi^\dagger \Phi^\dagger \Phi \Phi  \ dx \ dy}\eqno(2)$$ 
where the field of identical atoms each having mass $m$ is  represented by a complex scalar field operator $\Phi (\vec r, t)$. The potential $V (\vec r)$ 
 encapsulates
 interactions of atoms with electromagnetic fields from laser beams responsible for pinning the atoms to a two dimensional optical lattice.
  The second term
 in eq.(2) arises because
of binary contact interaction between atoms, $\lambda $ being a coupling constant  proportional to the s-wave scattering length.   

As a result of both the planar honeycomb lattice structure as well as the presence of a contact interaction term in eq.(2), the dynamics of 
 Bose-Einstein condensate (BEC) in 
this case is
  described by a nonlinear Dirac equation,
$$ i \hbar {{\partial \psi}\over{\partial t}} = \hat H (\psi) \psi = - i \hbar c_s (\vec \sigma . \nabla ) \psi  + M (\psi) \psi \eqno(3)$$
where $c_s \approx \mbox {few}\  cm\ s^{-1} $ is an effective BEC sound speed in the planar lattice, and $\psi $ represents the state of BEC,
\[
\psi =
\left( {\begin{array}{cc}
 \psi _A (x,y,t) \\
 \psi _B (x,y,t) \\
 \end{array} } \right)
\]
with $\psi_A$ and $\psi_B$ now being scalar wavefunctions representing spatio-temporal and pseudospin degrees of freedom for the bosonic atoms,
 while  $M (\psi) $ embodies the nonlinearity arising out of s-wave scatterings,
\[
M (\psi) = U
\left( {\begin{array}{cc}
 |\psi _A|^2 & 0  \\
 0 & |\psi _B|^2  \\
 \end{array} } \right)
\] 

$U$ being the interaction energy [9].

From eq.(3), one can straight away obtain the conservation of probability current,
$${1\over{c}} {{\partial P}\over {\partial t}} + \nabla . \vec J = 0 \eqno(4)$$
where,
$$ P\equiv \psi ^\dagger \psi = \psi^*_A \psi_A + \psi^*_B  \psi_B $$
and,
$$\vec J \equiv c_s \psi ^\dagger \vec \sigma \psi $$ 
Because of eq.(4), one may normalize the BEC wavefunction,
 $$ \int { }\int {P} \ dx \ dy = \int { } \int {(|\psi _A|^2 + |\psi _B|^2)} \ dx \ dy=1  \eqno(5)$$

\section{II. Dispersion Relation, Inner product and Energy}
In order to study dispersion relation in  NLDE, we  
 substitute the following plane wave form in eq.(3),
\[
\psi =
\left( {\begin{array}{cc}
 C _A  e^{i (k_x x +k_y y - \omega t)}\\
 C _B  e^{i (k_x x +k_y y - \omega t)}\\
 \end{array} } \right)
\]
that entails,

$$(\omega - \frac {U} {\hbar} | C_B |^2) C_B = c_s (k_x + i k_y) C_A \eqno(6a)$$
and,
$$(\omega - \frac {U} {\hbar} | C_A |^2) C_A = c_s (k_x - i k_y) C_B \eqno(6b)$$

Assuming box-normalization for $\psi $ on a square of side length $L$, one obtains from eq.(5),
$$ |C_A|^2 + |C_B|^2 = \frac {1} {L^2}\eqno(7)$$
Making use of eqs. (6a), (6b) and (7), we arrive at,
$$| C_A |^2 = \frac {1} {2L^2} \bigg(1 \pm \sqrt {1 - \frac {4 L^2} {U} \bigg (\hbar \omega - \hbar ^2 \Delta^2 \frac {L^2} {U}} \bigg) \bigg )\eqno(8)$$
where,
$$\Delta^2 \equiv \omega^2 - c^2_s (k^2_x + k^2_y) .\eqno(9)$$
Reality of $| C_A |^2$ requires,
$$ \frac {\hbar \omega} {U} \leq \frac {1} {4 L^2}  +  \frac {\hbar^2 L ^2 \Delta^2} {U^2} \eqno(10)$$
Therefore, for $U > 0$, eq.(10) implies that if,
$$ \hbar \omega \leq \frac {U} {4 L^2} + \frac {\hbar^2 L ^2 \Delta^2} {U} \eqno(11)$$
 plane wave solutions of eq.(3), corresponding to `massive modes' with tiny mass $\sqrt{\frac {\hbar^2 \Delta^2} {c^4}}$, are possible. 

If the inner product of two BEC states $\psi $ and $\chi $ is defined in the usual manner as,
$$(\psi , \chi) =\int { }\int {\psi ^\dagger \chi }\  dx \ dy \eqno(12)$$
one can trivially verify that,
$$(M (\psi) \chi, \psi)= (\chi, M(\psi) \psi)\eqno(13)$$
which entails,
$$(\hat H (\psi) \chi , \psi)= (\chi, \hat H (\psi) \psi) \eqno(14)$$
since $- i \hbar  (\vec \sigma . \nabla )$ is self-adjoint.

Therefore, if $\psi _E$ is an energy eigenstate so that,
$$\hat H(\psi _E) \psi _E = E \psi _E \eqno(15)$$
then substituting $\psi _E$ in place of $\chi $ and $\psi $ in eq.(14), leads to the result that E is real. However, if 
$$\hat H(\psi _i) \psi _i = E_i \psi _i \ , \ \ \  i=1,2$$
then, in general, $\psi_1 $ and $ \psi_2$ are not mutually orthogonal for non-zero $U$. 

(It must be pointed out here that technically we
should not be using terms like eigenvalues and eigenstates since we are dealing with a nonlinear operator $\hat H (\psi)$. But since
these terms have entered the common parlance of quantum theory, we will continue to use them.)

Suppose $U=0$ so that the Hamiltonian appearing in eq.(3) is just $- i \hbar c_s (\vec \sigma . \nabla )$. Then, it is interesting to observe 
that the orbital angular momentum normal to the planar lattice does not commute with the Hamiltonian,
$$[\hat L_z , -i\hbar \vec \sigma . \nabla ] = \hbar ^2 \bigg ( \sigma_x {{\partial}\over{\partial y}}
 - \sigma_y {{\partial}\over{\partial x}} \bigg ) \eqno (16)$$
Now, since,
$$[\sigma_z , -i\hbar \vec \sigma . \nabla ] = - 2 \hbar \bigg ( \sigma_x {{\partial}\over{\partial y}}
 - \sigma_y {{\partial}\over{\partial x}} \bigg ) \eqno (17)$$
we can get using eqs.(16) and (17), 
$$[L_z + {{\hbar}\over {2}} \sigma _z , -i\hbar \vec \sigma . \nabla ] = 0 .\eqno(18)$$
suggesting that $L_z + {{\hbar}\over {2}} \sigma _z $ plays the role of total angular momentum. But this is puzzling as $\vec \sigma $ appearing
 in eq.(3) reflects  only pseudospin degrees of freedom. In 
the context of graphene, using much more extensive arguments, Mecklenburg and Regan have recently made a case for identifying pseudospin with intrinsic
 spin angular momentum [10].
Although eq.(3) describes a bosonic system, the pseudospin degrees arising out of sites A and B of a two dimensional honeycomb
 lattice appear to induce a half-integral spin angular momentum for the BEC when $U=0$.
\section{III. Localized stationary solutions}
In this section, we look for energy eigenfunctions of the NLDE whose magnitudes fall  with radial distance. It is convenient to switch to
 polar coordinates $(r,\phi)$,
so that a stationary state solution $\psi _E (r,\phi, t)$ of the NLDE has the form,
$$\psi_E = e^{-{{i}\over{ \hbar}} E t} \psi (r, \phi)$$
where,\
\[
\psi (r, \phi)=
\left( {\begin{array}{cc}
 \psi _A (r, \phi) \\
 \psi _B (r, \phi) \\
 \end{array} } \right)
\]
satisfies the time independent NLDE,
$$E \psi_A = - i \hbar c_s e^{-i \phi} \bigg (\frac {\partial \psi_B} {\partial r} - \frac {i} {r} \frac {\partial \psi_B} {\partial \phi} \bigg ) + U |\psi_A|^2 \psi_A \eqno(19)$$
$$E \psi_B = - i \hbar c_s e^{i \phi} \bigg (\frac {\partial \psi_A} {\partial r} + \frac {i} {r} \frac {\partial \psi_A} {\partial \phi}\bigg ) + U |\psi_B|^2 \psi_B \ ,\eqno(20)$$

$E$ being the energy.

In polar coordinates, the orbital angular momentum is simply $\hat L_z = -i \hbar \frac {\partial} {\partial \phi} $. Therefore,
$$[\hat L_z ,  M (\psi)] = 0 $$
only when both $|\psi_A|^2$ and $|\psi_B|^2$ are independent of the angle $\phi $.
Since $[\sigma_z ,  M (\psi)] $ vanishes for all $\psi $, in order that $\hat L_z + \frac {\hbar} {2} \sigma_z $ commutes with $\hat H (\psi) $, 
 the two components of $\psi $ must take the following form,
$$\psi _A (r, \phi) = R_A (r) e^{i\Theta_A (r, \phi)} \eqno(21)$$
and,
$$\psi _B (r, \phi) = R_B (r) e^{i\Theta_B (r, \phi)} \eqno(22)$$
We seek solutions that are simultaneous eigenfunctions of the nonlinear Hamiltonian and the total angular momentum normal to the optical lattice.
So, making use of eqs.(21) and (22) in eqs.(19) and (20)
we find,
$$\Theta_A (r, \phi) = (l - \frac {1} {2}) \phi + f_A (r)\eqno(23)$$
$$\Theta_B (r, \phi) = (l + \frac {1} {2}) \phi + f_B (r)\eqno(24)$$
for $l=0,1,2 ...$, while the radial parts satisfy,
$$E R_A = - i \hbar c_s e^{i(f_B - f_A)} \bigg (\frac {d R_B} {d r} + i R_B \frac { d f_B} {d r} + \frac {R_B} {r} (l + \frac {1} {2}) \bigg ) + U R_A^3 \eqno(25)$$
$$E R_B = - i \hbar c_s e^{-i(f_B - f_A)} \bigg (\frac {d R_A} {d r} + i R_A \frac { d f_A} {d r} - \frac {R_A} {r} (l - \frac {1} {2}) \bigg ) + U R_B^3 \eqno(26)$$

When $R_A = R_B$, using the reality of energy $E$, one can deduce from eqs. (25) and (26)  the following class of exact solutions,
$$R_A = R_B = K r^{-1/2} \eqno(27)$$
$$f_B (r)= f_A(r) + Q\eqno(28)$$
where,
$$Q \equiv \sin ^{-1} {(- \frac {l \hbar c_s} {U K^2})}, \eqno(29)$$

with $K $ being the normalization constant.

If $l=0$, we get,
$$f_A(r)=\frac {(-1)^n} {\hbar c_s} [E r - U K^2 \ln {r}] + K_1\eqno(30)$$
and,
$$f_B (r)= f_A (r) + n \pi \eqno(31)$$
where $K_1 $ is an arbitrary real constant.

For other integral values of $l$, eqs. (25)-(27) demand  vanishing of energy eigenvalue along with,
$$f_A (r) = - \frac {U K^2} {\hbar c_s} \cos {Q} \ln {r} + K_2 ,\eqno(31)$$

 where $Q$ is given by eq.(29) and  $K_2 $ is a real constant.

If we normalize the wavefunction over a circular region of radius $L$, the  constant $K$ gets fixed to,

$$K=\frac {1} {\sqrt {4 \pi L}}$$

To summarize our results, we have obtained exact solutions of a nonlinear Dirac equation given by eq.(3) for the case $|\psi_A| = |\psi_B|$
that correspond to definite energy and angular momentum normal to the plane of the optical lattice. The position probability density for these
solutions decrease as $r^{-1}$. While the s-wave solution can have any value of
energy, solutions with $l=1,2..$ are static in nature since $E=0$.

What is interesting is that the planar honeycomb lattice not only leads to a massless nonlinear Dirac equation description of the BEC but also  makes
the pseudospin degrees of freedom appear as half-integral spin degrees in a purely bosonic system (also see [10]).
{\bf Acknowledgments}
We thank Sourin Das for stimulating discussions on the subject of graphene.

{\bf References}

[1] Novoselov, K. S. et al. {\it Science }{\bf 306}, 666-669 (2004).

[2] Novoselov, K. S. et al. {\it Proc. Natl Acad. Sci. USA} {\bf 102}, 10451-10453 (2005).

[3] Geim, A. K. and  Novoselov, K. S. {\it Nature Mat.} {\bf 6}, 183-191 (2007), and the references therein.

[4] Meyer, J. C. et al. {\it Nature }{\bf 446}, 60-63 (2007), and the references therein.

[5] Slonczewsky, J. C. and  Weiss, P. R. {\it Phys. Rev. }{\bf 109}, 272 (1958).

[6] Semenoff, G. W. {\it Phys. Rev. Lett.} {\bf 53}, 2449-2452 (1984).

[7] Haldane, F. D. M. {\it Phys. Rev. Lett.} {\bf 61}, 2015-2018 (1988).

[8] Katsnelson, M. I., Novoselov, K. S. and  Geim, A. K. {\it Nature Phys.} {\bf  2}, 620-625 (2006).

[9] Haddad, L. H. and Carr, L. D., {\it Physica D: Nonlinear Phenomena} {\bf 238},  1413 (2009).

[10] Mecklenburg, M. and  Regan B. C., arXiv:1003.3715
\end{document}